\documentclass[reprint, amssymb, amsmath, aip,cha]{revtex4-1}
\usepackage[version=3]{mhchem}
\usepackage[T1]{fontenc}
\usepackage{amsmath}
\usepackage{amssymb}
\usepackage{units}
\usepackage{graphicx}
\usepackage{psfrag}
\usepackage{tabularx}
\usepackage{booktabs}
\usepackage{bigstrut}
\usepackage{epstopdf}
\usepackage[caption=false]{subfig}

\begin{document}

\title{From Chemistry to Functionality:\\
Trends for the Length Dependence of the Thermopower in Molecular Junctions.} 
\date{\today}
\author{Falco H\"user}
  \affiliation{Nano Science Center
  and Department of Chemistry\\
  University of Copenhagen, 2100 K{\o}benhavn {\O}, Denmark}
\author{Gemma C. Solomon}
\email{gsolomon@nano.ku.dk} \affiliation{Nano Science Center
  and Department of Chemistry\\
  University of Copenhagen, 2100 K{\o}benhavn {\O}, Denmark}

\begin{abstract}
  We present a systematic ab-initio study of the length dependence of the 
  thermopower in molecular junctions.
  The systems under consideration are small saturated and conjugated molecular 
  chains of varying length attached to gold electrodes via a number
  of different binding groups.
  Different scenarios are observed: linearly increasing and decreasing 
  thermopower as function of the chain length as well as positive and negative
  values for the contact thermopower. Also deviation from the linear behaviour
  is found. The trends can be explained by details of the transmission, in
  particular the presence, position and shape of resonances from gateway
  states. We find that these gateway states do not only determine the contact
  thermopower, but can also have a large influence on the length-dependence
  itself. This demonstrates that simple models for electron transport do not
  apply in general and that chemical trends are hard to predict.
  Furthermore, we discuss the limits of our approach based on Density 
  Functional Theory and compare to more sophisticated methods like self-energy 
  corrections and the GW theory.
\end{abstract}


\maketitle

The perfect material for potential application in efficient thermoelectrical 
devices should be small in size, stable and, most of all, highly tunable.
In these respects, organic molecules seem very promising candidates,
since chemistry offers a vast number of possibilities for designing their 
electrical and thermal properties.\cite{Mahan96, Humphrey05, Wang06,
Venkataraman06-1, Venkataraman07, Kiguchi10, Capozzi14, Strange15}
In particular, simple carbon-based molecular wires bridging two gold electrodes 
via selected binding groups can provide systematic insight into fundamental
electronic and thermal transport mechanisms.
\cite{Ho08, Solomon08, Solomon09, Hines10, Lee13}
Ideally, mapping out chemical trends would be a new and easy way for predicting
a molecule's function. However, only little is known about the interplay
between chemical structure and thermoelectric features like conductance or
thermopower. The large number of degrees of freedom in molecular junctions,
such as the junction geometry, conformation and the chemical structure, makes
it particularly challenging to establish general trends.

Typical binding groups for gold electrodes are thiols\cite{Xu03} and
amines.\cite{Venkataraman06} More recently, the thermopower of single molecules
linked via isocyanide groups\cite{Tan11} as well as through direct covalent
\ce{Au}-\ce{C} bonds\cite{Cheng11} has been measured.
Amines exhibit a well-defined weak electronic coupling due to bonding of the 
\ce{N} lone pair to an undercoordinated \ce{Au} atom, allowing for stable
measurements.
Strong coupling is given by thiol binding groups, where a single sulfur atom can
bind to up to three gold atoms.\cite{Kristensen08}
However, uncertainties in the contact geometry lead to large variations in the 
measured and calculated conductances.
A well-defined contact geometry is achieved through the direct $\sigma$-type 
\ce{Au}-\ce{C} bond to an undercoordinated gold atom. This also allows for
direct coupling of the electrodes into the carbon backbone of the molecule.

In the linear response regime, the thermopower (also called Seebeck 
coefficient), is defined as
\begin{equation}
S = - \left.\frac{\Delta V}{\Delta T}\right|_{I=0},
\label{eq:S0}
\end{equation}
where $\Delta V$ is the thermally induced voltage at the steady state zero 
current due to an applied temperature difference, $\Delta T$, across the
junction.

It has been widely stated that a positive thermopower corresponds to hole 
transport or tunnelling through the highest occupied molecular orbital
(HOMO) and a negative thermopower to electron transport or tunnelling
through the lowest unoccupied molecular orbital (LUMO).
The question, however, is how these orbitals of the molecule in the junction
compare to the situation of the molecule in the gas phase.
In the presence of metallic electrodes, the molecular orbitals (MOs) become
renormalized and broadened.\cite{Neaton06} Charge transfer can lead to partial
occupation numbers. Due to hybridization of localized MOs with metallic
states at the interface with the electrodes, so-called gateway states can
appear within the HOMO-LUMO gap.\cite{Widawsky12, Kim14, Khoo15}

Experimentally, it has been found that the thermopower increases linearly with 
the length of a molecular chain, $N$.\cite{Reddy07, Malen09}
This is supported in theory for the off-resonant tunnelling regime,
where the transmission is exponentially surpressed around the Fermi 
energy\cite{McConnell61, Karlström14}
\begin{equation}
\tau(E) = \alpha(E)e^{-\beta(E)N},
\label{eq:tau_exp}
\end{equation}
with parameters $\beta(E)$ and $\alpha(E)$, which determine the decay factor 
and the contact resistance, respectively.
Then, the thermopower becomes\cite{Pauly08}
\begin{equation}
S \propto \beta'(E)N - \left.\frac{\alpha'(E)}{\alpha(E)}\right|_{E=E_F}.
\label{eq:Slin}
\end{equation}
Here, $'$ denotes the derivative with respect to energy.
Eq.~(\ref{eq:Slin}) implies that the length dependence of the thermopower
(the slope) is given by $\beta_S = \beta'(E)|_{E=E_F}$,
whereas the contact thermopower, $S(N=0)$, only depends on $\alpha(E)$.

Recent tight-binding calculations support the common assumption that $\beta(E)$ 
is related to the molecular backbone and position of the MO energies,
since it is independent of the contact coupling.\cite{Karlström14}
$\alpha(E)$, on the other hand, is directly associated with states on the 
binding groups and has a strong dependence on the coupling strength.
Depending on the position of the Fermi level and the binding group energy, four 
different scenarios for the length dependence of the thermopower
can in principal be found: positive or negative contact thermopower with 
increasing or decreasing values with length.
A case, where the thermopower changes sign with increasing molecular length
has been measured recently for the first time for oxidized 
oligothiophenes.\cite{Dell15}

\section{Method\label{sec:method}}
The transmission, $\tau(E, V)$, is calculated within the 
Landauer-B\"{u}ttiker formalism for coherent transport\cite{Meir92}
using standard nonequilibrium Green's Functions techniques based on Density 
Functional Theory (NEGF-DFT).\cite{Taylor01, Xue02, Brandbyge02, Thygesen03}
Within this framework, the thermopower is given as
\begin{equation}
S = - \frac{1}{e T} \frac{\int \! dE \, E \, \tau(E, V=0) \, f'(E, T)}
{\int \! dE \, \tau(E, V=0) \, f'(E, T)},
\label{eq:S1}
\end{equation}
where $e$ is the elementary charge, $T$ the absolute temperature and $f(E, T)$ 
the Fermi distribution.
In the limit of small temperature gradients, $\Delta T \rightarrow 0$,
Eq.~(\ref{eq:S1}) reduces to\cite{Butcher90, Paulsson03, Segal05, Malen10}
\begin{equation}
S = - \frac{\pi^2 k_B^2 T}{3 e}
\left.\frac{\partial \ln(\tau(E), V=0)}{\partial E}\right|_{E=E_F},
\label{eq:S2}
\end{equation}
where $k_B$ is the Boltzman constant.
All thermopower values presented in this work have been calculated from
Eq.~(\ref{eq:S1}) at room temperature ($T = \unit[300]{K}$).
Eq.~(\ref{eq:S2}) illustrates that the thermopower is proportional to the 
negative logarithmic derivative of the transmission at the Fermi energy
of the electrodes, $E_F$.

All calculations have been carried out with the electronic structure code 
GPAW\cite{Enkovaara10} using the generalized gradient PBE
exchange-correlation functional.\cite{Perdew96} A double-zeta polarized basis 
set of diffuse basis functions with a
confinement-energy shift of $\unit[0.01]{eV}$ was employed.\cite{Strange11-1}
The junctions were modelled in supercells with 7-8 layers of gold containing 16 
atoms each arranged in the fcc structure. A $(4\times4)$ $k$-point sampling was 
applied in the directions of the electrode surface and the
transmission was averaged over all $k$ points.\cite{Thygesen05, Strange11-1}
The coupling to the bulk electrodes is obtained using standard 
techniques.\cite{Strange08, Strange11-1}

\begin{figure*}[t]
\begin{centering}
\includegraphics[width=\textwidth,clip=]{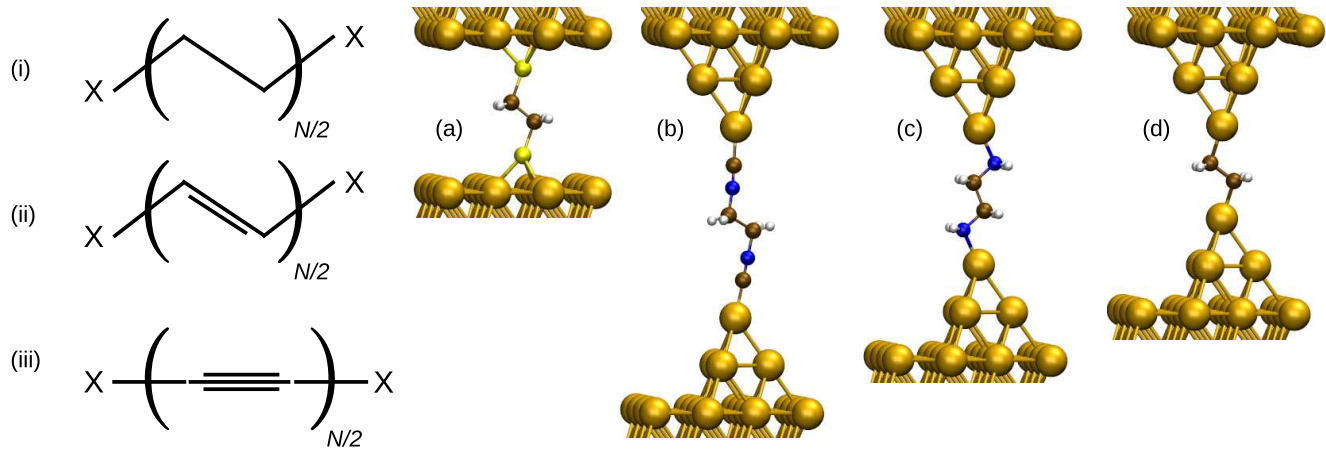}
\par\end{centering}
\caption{Left side:
Chemical structure of (i) alkane, (ii) alkene and (iii) alkyne chains.
\ce{X} = \ce{S}, \ce{NH2}, \ce{NC}, - (direct coupling).
$N$ is the number of \ce{(CH2)2}, \ce{(CH)2} and \ce{(C)2} units, respectively.
Right side:
Junction geometries for (a) thiol-, (b) isocyanide-
and (c) amine-end groups and (d) direct coupling.}
\label{fig:geometries}
\end{figure*}

The junction geometries are sketched in Fig.~\ref{fig:geometries}: Nitrogen and 
carbon atoms were chosen to bind to a single gold adatom
at the tip of a small pyramid, whereas sulfur atoms were placed in a fcc-hollow
site of a flat gold surface. For each binding group, alkanes
(saturated carbon chains), alkenes (conjugated carbon chains with alternating
single and double bonds) and alkynes (conjugated carbon chains with alternating
single and triple bonds) with 2, 4, 6 and 8 carbon units were modelled.
Geometry optimization was performed for all molecules
(including the tip atoms in the pyramid configuration) on ideal gold surfaces.

\section{Results\label{sec:results}}
\subsection{Alkane chains\label{subsec:alkanes}}
\begin{figure}
\begin{centering}
\includegraphics[width=\columnwidth,clip=]{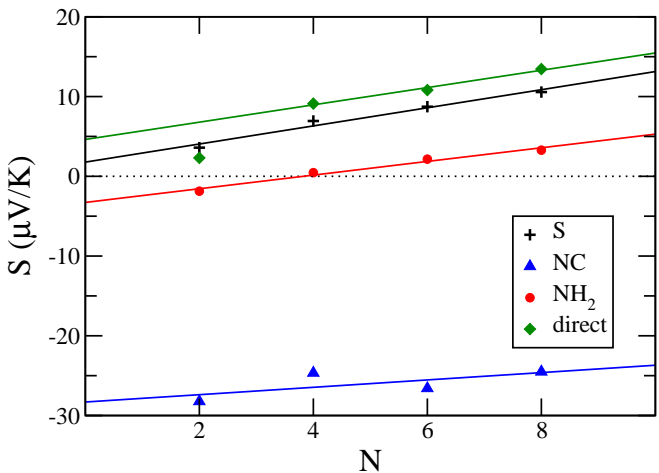}
\par\end{centering}
\caption{Calculated thermopower for alkane chains with amine-, isocyanide- and 
sulfur-binding groups as well as direct coupling in the junction geometries of
Fig.~\ref{fig:geometries}.
Straight lines are linear fits to $S(N) = S(0) + \beta_S \cdot N$,
where $N$ is the number of \ce{(CH2)2} units. All fitted values are listed in 
Table \ref{tab:values}.
The first data point is not considered in the fitting for direct coupling.
The dotted horizontal line serves as guide for the eye at $S = 0$.}
\label{fig:thermopower_alkanes}
\end{figure}

The thermopower for the alkane chains is shown in
Fig.~\ref{fig:thermopower_alkanes}. It increases roughly linearly with length
for all binding groups with a rather small prefactor of around
$\unit[0.5 - 1.1]{\mu V/K}$ per methyl unit. The onset, however, is very
different for the four binding groups considered here:
\ce{NH2} gives a small and \ce{NC} a large negative value, respectively,
whereas the onset is close to $0$ for sulfur and slightly positive for direct 
coupling.
This results in different scenarios: Positive and increasing thermopower for 
thiol binding groups as well as direct coupling,
and negative and increasing thermopower for isocyanide binding groups.
For the amine linkers, the thermopower changes sign
with increasing length, going from negative values for short chains to positive
values for longer chains ($\geq$ 4 methyl units).

\begin{figure}
\begin{centering}
\includegraphics[width=\columnwidth,clip=]{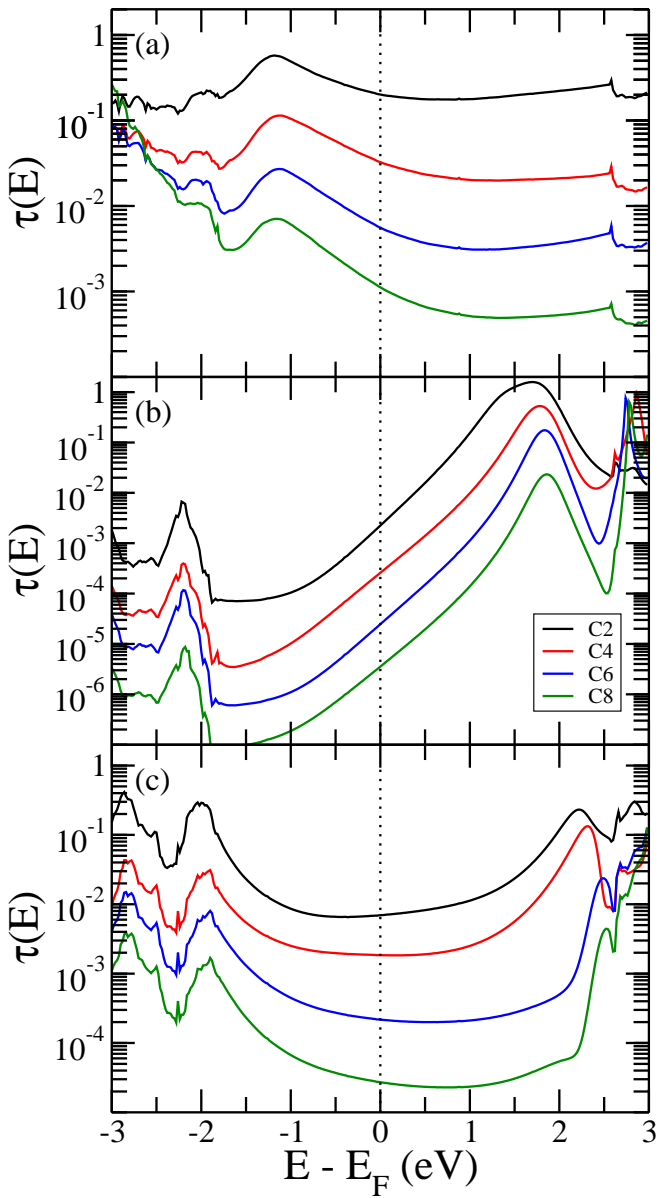}
\par\end{centering}
\caption{Transmission curves for alkane chains with (a) thiol, (b) isocyanide 
and (c) amine binding groups and 2, 4, 6 and 8 methyl units.
The dotted line indicates the Fermi level, which is set to $0$.}
\label{fig:TE_alkanes}
\end{figure}

The reasons for these different behaviours can be seen from the corresponding
transmission curves presented in Fig.~\ref{fig:TE_alkanes}:
For thiol end groups, a broad gateway state resonance originating from 
hybridization of states located on the sulfur atoms
with \ce{Au} d-states appears within the HOMO-LUMO gap about $\unit[1.2]{eV}$ 
below the Fermi level resulting in a negative slope of $\tau(E)$
and thus a positive thermopower. With increasing chain length, the gap becomes 
deeper without changing the shape or position of the gateway state
resonance. Since the center of the gap falls off faster than the resonance peak 
height, the slope of the transmission curve at $E_F$ becomes
steeper and the thermopower larger.

This situation is similar for the chains with direct coupling (see Supporting 
Information),
where the gateway resonance is much broader. For the shortest chain, however, 
the hybridization is so strong that the transmission
becomes very flat with values close to 1 around the Fermi energy. In this case, 
the transport mechanism is resonant tunnelling
and Eqs.~(\ref{eq:tau_exp}) and (\ref{eq:Slin}) do not apply.\cite{Cheng11}
Therefore, the first data point in Fig.~\ref{fig:thermopower_alkanes} is off 
the straight line.

For the isocyanides on the other hand, the transmission is dominated by a 
relatively sharp gateway state resonance
slightly above the Fermi energy giving positive slopes and large negative 
Seebeck coefficients.
As for the previous molecules, the peak height does not drop off as fast as the 
center of the gap for longer chains.
Additionally, the center of this peak moves slightly up in energy.
This results in an increase of the slope of the transmission curve and thus the 
thermopower.

In the case of amine-end groups, the Fermi level is located around the middle 
of the HOMO-LUMO gap,
where the transmission curve is very flat. This is why the obtained values for 
S are small and even subtle variations in the position
of a resonance imply a change of sign of the thermopower.

\begin{table}[t]
\begin{centering}
\caption{\label{tab:values}Fitted contact thermopower, $S(0)$, and length 
dependence, $\beta_S$.}
\begin{tabularx}{\columnwidth}{l | r | r}
\hline\hline
	        & $S(0) (\unit[]{\mu V/K})$& $\beta_S (\unit[]{\mu V/K})$\\
\hline
alkane-\ce{NH2} & $-3.3$	 	 & $0.86$ \\
alkane-\ce{NC}  & $-28.3$		 & $0.46$ \\
alkane-\ce{S}   & $1.8$			 & $1.14$ \\
alkane-direct   & $4.6$			 & $1.09$ \\
\hline
alkene-\ce{NH2} & --			 & --	  \\
alkene-\ce{NC}  & $-72.8$		 & $-8.60$\\
alkene-\ce{S}   & $3.0$			 & $2.53$ \\
alkene-direct   & $-5.3$		 & $3.51$ \\
\hline
alkyne-\ce{NH2} & --			 & --	  \\
alkyne-\ce{NC}  & $-44.9$		 & $-6.91$\\
alkyne-\ce{S}   & $7.6$			 & $1.51$ \\
alkyne-direct   & $-5.5$		 & $5.85$ \\
\hline\hline
\end{tabularx}
\end{centering}
\end{table}

\subsection{Alkene chains\label{subsec:alkenes}}
\begin{figure}
\begin{centering}
\includegraphics[width=\columnwidth,clip=]{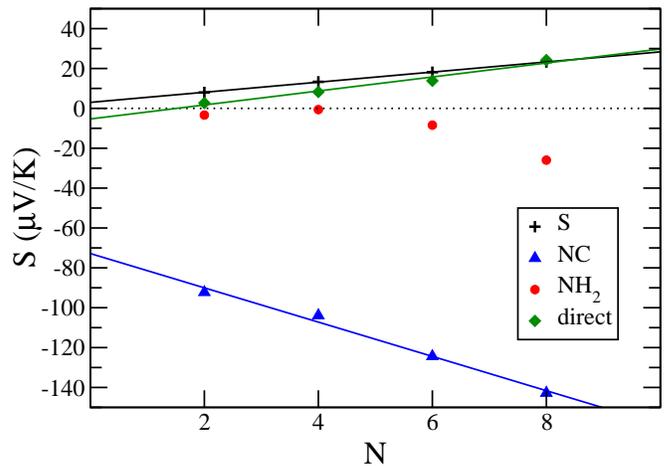}
\par\end{centering}
\caption{Same as Fig. \ref{fig:thermopower_alkanes} for alkene chains.
For the amine-end groups, no linear fitting could be performed.}
\label{fig:thermopower_alkenes}
\end{figure}

The results for alkene chains are plotted in Fig.~\ref{fig:thermopower_alkenes}.
Again, the calculated values for the thermopower for the amine-end groups are 
very small, but we cannot see a linear dependence with length. In the
corresponding transmission curves (see SI), a sharp very resonance appears at
around $\unit[0.5]{eV}$ above the Fermi energy,
which arises from the LUMO of the molecule in the gas phase.
This orbital is localized on the carbon backbone of the molecule and on the 
hydrogen atoms of the amine groups that stick out in the direction of the
$\pi$-system of the chain, thereby forming a hyperconjugated state.
There is no contribution on the nitrogen atoms.
Therefore, this state does not couple to the gold electrodes when the molecule 
is put into the junction in the given contact geometry.
This situation makes it impossible to distinguish between backbone and end 
group effects on the transmission properties.
A possible workaround would be to rotate the hydrogen atoms out of the 
$\pi$-system of the backbone.
However, we could not find an energetically stable conformation or junction 
geometry for that.

\begin{figure}
\begin{centering}
\includegraphics[width=\columnwidth,clip=]{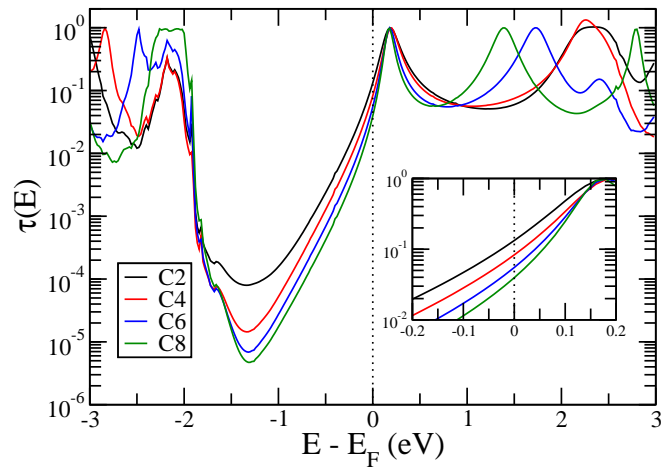}
\par\end{centering}
\caption{Transmission for alkene-\ce{NC} chains. Inset: Zoom in around the 
Fermi energy.}
\label{fig:TE_alkene-NC}
\end{figure}

A new situation occurs for the isocyanide end-groups:
the thermopower has a large negative onset and decreases even more with length.
As shown in Fig.~\ref{fig:TE_alkene-NC}, there is a sharp resonance from the 
LUMO slightly above $E_F$ reaching a transmission value of $1$.
It does neither decrease nor shift in energy for longer chains.
Therefore, as the transmission gap deepens, the slope of the transmission
becomes steeper and the thermopower decreases.

For the thiol-end groups, the thermopower is again positive and increasing as 
for the alkane chains, but with larger values for the onset and slope of
$S(N)$.
A small negative contact thermopower is found for the directly coupled chains,
although the thermopower itself is positive for all molecules and increases 
with length.
As opposed to the alkanes, we find a large spread of values for $\beta_S$ with 
both negative and positive values.

\subsection{Alkyne chains\label{subsec:alkynes}}
\begin{figure}
\begin{centering}
\includegraphics[width=\columnwidth,clip=]{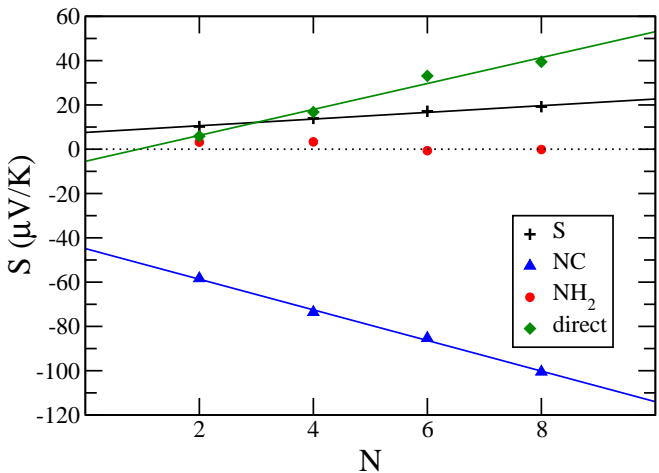}
\par\end{centering}
\caption{Same as Fig.~\ref{fig:thermopower_alkanes} for alkyne chains.
For the amine-end groups, no linear fitting could be performed.}
\label{fig:thermopower_alkynes}
\end{figure}

The results for alkyne chains are presented in
Fig.~\ref{fig:thermopower_alkynes} and are similar to the alkenes.
There are four different scenarios for the length dependence of the thermopower:
positive and increasing (\ce{S}), negative and decreasing (\ce{NC}), negative 
and increasing (direct) and changing sign from positive to negative (\ce{NH2}).

All further transmission curves are given in the Supporting Information.
The fitted values for the contact thermopower and the slope are listed in Table 
\ref{tab:values}.

\section{Discussion\label{sec:discussion}}
\subsection{Contact thermopower\label{subsec:contactS}}
By comparing all of our results, we do see a trend for the contact thermopower, 
$S(0)$, for the different binding groups:
For thiols, it is positive for all systems studied here. This is in agreement 
with experimental findings for alkanedithiols\cite{Malen09, Guo13} and
phenyldithiols.\cite{Reddy07, Tan10, Tan11, Balachandran13}
Large negative values are obtained in all cases for isocyanide binding groups.
This observation matches with measurements for \ce{NC}-terminated 
phenyls.\cite{Tan11}

The situation is not entirely clear for the directly coupled chains:
While all calculated values for the thermopower itself were positive,
we find both positive and negative contact thermopowers in the linear fitting.
This is a result from the appearance of very broad gateway resonance close to 
the Fermi energy, as mentioned in the previous section for alkane chains.
For short lengths, resonant rather than off-resonant tunneling occurs.

The contact thermopower is expected to be around $0$ for amine-terminated 
chains, which would agrees with measurements on phenyldiamines\cite{Malen09}
but no linear length dependence could be found for alkenes and alkynes.
The calculated values for the thermopower itself, however, are all very close
to $0$.

\subsection{Length dependence\label{subsec:slopeS}}
The slope, $\beta_S$ does not follow a simple rule of thumb, since neither its 
value nor even its sign can be directly predicted from the properties of the
backbone alone. In particular, we find a large variation for the
conjugated molecules. However, this lies within the predictions of the 
tunneling model as described in Ref.~\citenum{Karlström14}:
$\beta(E)$ exhibits a strong energy dependence and its derivative, $\beta'(E)$, 
changes sign around the Fermi level.
Since end groups can strongly influence the position of MO energies with
respect to $E_F$ (alteration of the band lineup or level alignment)
\cite{Xue04, Zotti10, Balachandran13} they also effect
$\beta_S = \beta'(E)|_{E=E_F}$. The length dependence of the thermopower is
thus not only a property of the backbone, as previously assumed, but rather of
the backbone + binding groups. In the presence of gateway states, 
it even depends on the whole junction.

Early measurements reported positive and decreasing thermopower with length for 
alkanedithiols\cite{Malen09}, whereas positive and increasing
values were measured more recently\cite{Guo13}, the latter being consistent 
with our calculations.
We note that we have also calculated the thermopower for \ce{S}-terminated 
chains in the top-geometry (as for the other binding groups).
While the transmission curves look very different (see SI), we still find a 
positive and increasing thermopower.
However, our findings suggest that the length dependence can be very sensitive 
to details in the coupling,
which might explain the differences in the two experiments.

In the experiments of Ref.~\citenum{Widawsky13}, a positive and increasing 
thermopower was found
for directly \ce{Au}-\ce{C} coupled alkane chains.
The dependence was nonlinear and a saturation of $S$ with molecular length was 
observed for a length of 10 methyl units.
However, this data set contained only three points. Thus, we don't see a 
contradiction to our results.

\subsection{Methodology\label{subsec:DFT}}
It is well known that DFT tends to overestimate conductances by up to several 
orders of magnitude and that higher levels of theory, such as the GW method,
are required to bring calculated values in good agreement with experimental
results.\cite{Strange11, Strange11-1, Jin13, Markussen13}
Even though the thermopower is most likely a more robust quantity, since it
is found to be less sensitive to errors in method,\cite{Ke09} DFT cannot be
expected to give quantitatively accurate results.
As discussed here, the thermopower depends mainly on the presence, position and 
shape of resonances from gateway states.
DFT usually places molecular resonances too close to the Fermi level, leading
to large errors in transport calculations. This is a consequence of the
self-interaction error inherent in most functionals, which results in an
inaccurate molecule-lead charge transfer.\cite{Ke07}
Another problem is the underestimation of the HOMO-LUMO gap with DFT, resulting
in far too low values for the conductance.
Nevertheless, it is possible to predict qualitative trends. In particular, 
correct signs of the thermopower and the linear prefactor, $\beta_S$, have been
found.\cite{Ke09, Quek11, Widawsky12}

Good results for the conduction and thermopower could also been obtained within 
the DFT~+~$\Sigma$ approach, which corrects for static correlation and image
charge effects.\cite{Quek09-1, Quek11, Widawsky13, Kim14}
This method shifts all occupied (unoccupied) molecular orbitals down (up)
in energy by a constant value. Thus, it moves the molecular resonances
further away from the Fermi level, which results in a deepening of the
transmission gap. It is, however, not clear how gateway states are affected.
In any case, DFT~+~$\Sigma$ fails for systems where charge transfer from the 
electrodes leads to partial occupation of otherwise unoccupied molecular
orbitals.\cite{Khoo15} This turned out to be the case for many of the systems
in our study. A detailed comparison is given in the Supporting Information.

Much better results are expected for the GW method. This, however, is 
computationally extremely challenging and we were not able to
converge results properly with respect to the basis set and the size of the 
interacting region for many of the structures.

\section{Conclusions\label{sec:conclusions}}

We have calculated the thermopower of small molecular chains attached to gold 
electrodes in well-defined and consistent binding geometries using Density
Functional Theory.
By varying the degree of conjugation and binding group, we have investigated
the question to what extent their chemical properties determine the length 
dependence of the thermopower and if predictions can be made.
A simple assumption that the thermopower depends linearly on the length and 
that this dependence is given by the nature of the molecular backbone does not
hold. Instead, we find that in many cases resonances from gateway states
dominate the electron transmission close to the Fermi level of the electrodes 
and thus the transport properties of the molecule.
It is therefore the binding group, that not only dictates the contact 
thermopower, but also influences the length dependence to a large degree.
A situation, where the gateway states are completely decoupled from the 
backbone in order to seperate their effects would be desirable,
but could not be realized with reasonable junction geometries for many of the 
short molecules. The importance of gateway resonances is diminished for longer
chains. However, we expect their influence to be neglibile only for much
greater lengths than studied here. Overall, we can identify different scenarios
for the length dependence. Amongst others, some where the thermopower changes
sign with length.

Although DFT is not expected to yield quantitatively accurate results, we 
believe that it gives correct trends. Including static self-energy corrections
in the so-called DFT~+~$\Sigma$ approach did not prove to be feasible in this
broad study, since its validity is restricted to cases where interactions
between molecule and electrode surface are sufficiently low.
Unfortunately, the self-consistent GW method did not turn out to be generally
applicable either, as many of the calculations were practically impossible
to converge.

In summary, we have demonstrated that, although it is very difficult to
separate end group from backbone effects, the thermopower can be tuned
chemically, even for the simplest systems. Other chemical modifications
may lead to further variations in the values for the thermopower. However,
the influence from the binding groups cannot be neglected for short molecules.

\section{Acknowledgements}
This project has received fundings from the European Research Council
under the European Union's (EU) Seventh Framework Program (FP7/2007-2013)/ERC 
Grant Agreement No. 258806.

\section{Supplementary Information}
Transmission curves obtained with DFT and DFT~+~$\Sigma$ for all 60 molecules
(including an alternative binding geometry for thiols).

\bibliography{bibfile}{}

\end{document}